\documentclass[12pt]{article}
\usepackage{a4wide}

\newcommand{\be}{\begin{equation}}
\newcommand{\ee}{\end{equation}}
\newcommand{\ba}{\begin{eqnarray}}
\newcommand{\ea}{\end{eqnarray}}
\newcommand{\ve}{\varepsilon}
\newcommand{\lgl}{\langle}
\newcommand{\rgl}{\rangle}
\newcommand{\epe}{\epsilon^{\mu\nu\alpha\beta}}

\newcommand{\um}{u_\mu}
\newcommand{\un}{u_\nu}
\newcommand{\ua}{u_\alpha}
\newcommand{\ub}{u_\beta}
\newcommand{\ug}{u_\gamma}

\newcommand{\cm}{\chi_-}
\newcommand{\cp}{\chi_+}
\newcommand{\fmmn}{f_{-\mu\nu}}
\newcommand{\fmna}{f_{-\nu\alpha}}
\newcommand{\fmab}{f_{-\alpha\beta}}
\newcommand{\fmgm}{f_{-\gamma\mu}}
\newcommand{\fmgn}{f_{-\gamma\nu}}

\newcommand{\fmgb}{f_{-\gamma\beta}}

\newcommand{\fpmn}{f_{+\mu\nu}}
\newcommand{\fpna}{f_{+\nu\alpha}}
\newcommand{\fpab}{f_{+\alpha\beta}}
\newcommand{\fpgm}{f_{+\gamma\mu}}
\newcommand{\fpgn}{f_{+\gamma\nu}}
\newcommand{\fpga}{f_{+\gamma\alpha}}

\newcommand{\hmn}{h_{\mu\nu}}
\newcommand{\hgm}{h_{\gamma\mu}}
\newcommand{\hgn}{h_{\gamma\nu}}

\newcommand{\hgb}{h_{\gamma\beta}}

\newcommand{\trace}[1]{\langle #1 \rangle}
\newcommand{\com}[1]{\left[ #1 \right]}
\newcommand{\acom}[1]{\left\{ #1 \right\}}

\begin{document}
\renewcommand{\thefootnote}{\fnsymbol{footnote}}

\setcounter{equation}{0}
\setcounter{subsection}{0}
\setcounter{table}{0}
\setcounter{figure}{0}

\begin{titlepage}
\begin{flushright}
LU TP 01-34\\
DFPD-01/TH/25\\
CPT-2001/P.4256\\
hep-ph/0110400\\
\end{flushright}
\begin{center}
\begin{bf}
{\Large \bf 
The anomalous chiral Lagrangian of order~$p^6$
} \\[2cm]
\end{bf}
{\large J. Bijnens$^{a}$, L. Girlanda$^{b}$ and P. Talavera$^{c}$}
\\[1cm]
$^a$ 
Dept. of Theor. Phys. 2, Lund University,\\
S\"olvegatan 14A,
S-22362 Lund, Sweden\\[.1cm]
$^b$   
Dipartimento di Fisica, Universit\`a di Padova and
INFN,\\
Via Marzolo 8, I-35131 Padova, Italy\\[.1cm]
$^c$   
Centre de Physique Th\'eorique, CNRS--Luminy, Case 907\\
F-13288 Marseille Cedex 9, France\\[.1cm]
\end{center}

\vfill

\begin{center}
{\bf Keywords:} \begin{minipage}[t]{9.5cm} 
Chiral Lagrangians, Nonperturbative Effects,\\ Spontaneous Symmetry Breaking, QCD.
\end{minipage}
\end{center}

\vfill
\begin{abstract}
We construct the effective
chiral Lagrangian for chiral perturbation theory
in the mesonic odd-intrinsic-parity sector at order $p^6$. The
Lagrangian contains 24 in principle measurable terms and no contact terms 
for the general case
of $N_f$ light flavours, 23 terms for three and 5 for two flavours.
In the two flavour case we need a total of 13 terms if an external
singlet vector field is included. We discuss and implement the methods
used to reduce to a minimal set. The infinite parts needed for
renormalization are calculated and presented as well.
\end{abstract}
\vfill
\end{titlepage}

\setcounter{footnote}{0}
\renewcommand{\thefootnote}{\arabic{footnote}}

\setcounter{page}{1}
\setcounter{equation}{0}
\section{\bf Motivation}

Effective field theory methods are widely used in physics.
Many rely on the spontaneous breaking of an internal continuous 
symmetry, represented  by a compact connected
Lie group, G, to a subgroup, H. In the breakdown ``massless excitations''
appear,  
which are usually referred to as Goldstone boson modes,
$\pi^a(x)$. They parametrize the coset space G/H in terms of a general
spacetime dependent G transformation ${\cal U}(\pi)$  that 
transforms non-linearly under G and linearly under H. 
To be more definite, under a global transformation $g\in G$ one maps
$\pi \rightarrow \pi^\prime$ via
\begin{equation}
g {\cal U}(\pi) \rightarrow {\cal U}(\pi^\prime) h(\pi,g)\,,
\end{equation}
where $h(\pi,g)$ is an element of the unbroken subgroup H \cite{Weinberg}. 
In order to parametrize the low-energy dynamics of a physical system
one constructs the most general G-invariant 
Lagrangian density as a sum of monomials, which is an invariant product of
covariant derivatives 
defined on the subalgebra corresponding to H, ${\cal H}$ \cite{Weinberg}
\begin{equation}
\label{eff}
{\cal L}({\cal U}^{-1} D_\mu^{\cal H} {\cal U}, \ldots)\,.
\end{equation}
External fields and explicit symmetry breaking effects can be
included as well.
However this procedure does not lead to the most general G-invariant action S. 
As is well-known \cite{WZ}, operators in the Lagrangian density that
are not invariant under G can lead
to a G-invariant action if their variation under G is a total derivative.
These operators form the anomalous sector and they  
arise already at the classical level in the effective
field theory.

The power of an effective field theory as given in (\ref{eff}) 
is that its construction only relies on the symmetries
of the initial group, broken subgroup and some parameter [derivative]
expansion. The latter, 
with more or less phenomenological insight, is related to the physical problem
under study. It is thus clear 
that the success of the approach depends on the construction of the complete
string 
of operators which form the Lagrangian. The general
recipe for the construction is based on writing all possible terms
allowed by the continuous and discrete
symmetries of the physical problem \cite{Weinberg2}. 
The omission of any term would lead to an inconsistent parametrization of
the low-energy dynamics invalidating the effective approach.
On the other side a parametrization with redundant operators will make
the identification of the 
relevant operators in a physical problem difficult.
Therefore it is convenient to obtain the minimal structure of the Lagrangian
density.  
Since a long time ago it is known that the anomalous Lagrangian 
has a clear and nice
geometrical interpretation \cite{Witten}. For QCD-like theories,
at leading order in four spacetime dimensions
and in the absence of external
currents, it is constrained to contain a single term \cite{D'Hoker} fixed
entirely by the single generator of the fifth de Rham
cohomology group \cite{Manes:1985df}. 
Several classifications of the relevant operators at next-to-leading order 
have been performed \cite{Issler2,Akhoury,FearingScherer2}, but no
agreement was reached regarding the number of independent operators
involved\footnote{While we were preparing this manuscript Ref. \cite{Fearing2}
appeared using a different basis.}.
We clarify this issue, in particular supplementing
the previous analyses with a geometrical missing ingredient: the Bianchi
identities.  Together with the
construction of  
a minimal basis at next-to-leading order we provide the divergent part needed
to cancel the ultra-violet behaviour  
of the loop graphs. This together with the full list and infinities
for the even-intrinsic-parity sector of \cite{BCGlag} completes
the construction of the full $p^6$ mesonic Lagrangian.
We use here the standard power counting where quark masses
count as $p^2$ \cite{GL}. For an alternative counting see e.g. \cite{GCHPT}.

\setcounter{equation}{0}
\section{Leading order Lagrangian construction}

We shall focus the analysis entirely on 4-dimensional field theories where
the free 
gauge group G is coupled vectorially to $N_f$ massless Dirac fermions. All
of them transform according to some 
irreducible representation $r$ of the gauge group. 
Under the assumption of maximal breaking of chiral symmetry with
simultaneous preservation 
of maximal flavour symmetry there are three allowed scenarios of spontaneous
symmetry breaking  
\cite{Peskin:1980gc} depending on the representation $r$. 
We shall assume that the pattern is given by \cite{vafawitten}
\begin{equation}
\label{ssb}
SU(N_f)_L \times SU(N_f)_R \rightarrow SU(N_f)_V\,.
\end{equation}

The leading-order anomalous Lagrangian in the absence of external sources is
reduced to only a single term 
\cite{Witten}.
In the case of a spontaneous symmetry breaking given by  (\ref{ssb}) the
homotopy group is trivial 
and a smooth interpolating field between the 4-dimensional Minkowski
spacetime and a 5-dimensional 
ball, $B_5$, can be found, allowing then to write the action as
\begin{equation}
\label{ball}
s[x] = \int_{B_5} \,d^4x\,dt\,{\cal L}_1\,,
\end{equation}
where
${\cal L}_1$ is a G-invariant density and its form is restricted by
integrability conditions.  
Using Poincar\'e's lemma, (\ref{ball}) can be cast in terms of a closed
5-form on $G/H$. The  
terms that cannot be reduced to four-dimensional integrals are directly
obtained by the generators 
of the fifth de Rham cohomology group $H^5(G/H;{\bf R})$. In the case where
the coset subgroup 
is a simple Lie group, $H^5$ has a single generator
\begin{equation}
\Omega_5 = \frac{-i}{240 \pi^2} \langle( U^{-1} dU )^5\rangle\,,
\end{equation}
which is the Wess-Zumino-Witten term.
$\langle A\rangle$ corresponds to the trace over the matrix $A$.

In order to describe phenomenologically relevant processes,
the Goldstone boson modes need to be coupled to external gauge invariant
sources. This obviously 
increases the number of 
possible structures. All the new structures are reduced to exterior
derivatives of an invariant 
4-form. Such  exterior derivatives in terms of the initial 5-form
yield terms in 
(\ref{ball}) that can be written as four-dimensional integrals of a
G-invariant density. 
Explicitly the full action at next-to-leading order can be written as
\cite{Witten}
\ba
\lefteqn{
S[U,\ell,r]_{\mathrm{WZW}}  =  -\,\displaystyle \frac{i N_C}{240 \pi^2}
\int d\sigma^{ijklm} \left\langle \Sigma^L_i
\Sigma^L_j \Sigma^L_k \Sigma^L_l \Sigma^L_m \right\rangle }
\nonumber\\ &&\mbox{}
 -\,\displaystyle \frac{i N_C}{48 \pi^2} \int d^4 x\,
\varepsilon_{\mu \nu \alpha \beta}\left( W (U,\ell,r)^{\mu \nu
\alpha \beta} - W ({\bf 1},\ell,r)^{\mu \nu \alpha \beta} \right) ,
\label{eq:WZW}
\ea
\ba
\lefteqn{
W (U,\ell,r)_{\mu \nu \alpha \beta}  = 
\left\langle U \ell_{\mu} \ell_{\nu} \ell_{\alpha}U^{\dagger} r_{\beta}
+ \displaystyle \frac{1}{4} U \ell_{\mu} U^{\dagger} r_{\nu} U \ell_\alpha U^{\dagger} 
r_{\beta}
\right. }
\nonumber\\ && \mbox{}
+ i U \partial_{\mu} \ell_{\nu} \ell_{\alpha} U^{\dagger} r_{\beta}
 +  i \partial_{\mu} r_{\nu} U \ell_{\alpha} U^{\dagger} r_{\beta}
- i \Sigma^L_{\mu} \ell_{\nu} U^{\dagger} r_{\alpha} U \ell_{\beta}
\nonumber\\ && \mbox{}
+ \Sigma^L_{\mu} U^{\dagger} \partial_{\nu} r_{\alpha} U \ell_\beta
 -  \Sigma^L_{\mu} \Sigma^L_{\nu} U^{\dagger} r_{\alpha} U \ell_{\beta}
+ \Sigma^L_{\mu} \ell_{\nu} \partial_{\alpha} \ell_{\beta}
+ \Sigma^L_{\mu} \partial_{\nu} \ell_{\alpha} \ell_{\beta}
\nonumber\\ &&  \left.\mbox{} 
 - i \Sigma^L_{\mu} \ell_{\nu} \ell_{\alpha} \ell_{\beta}
+ \displaystyle \frac{1}{2} \Sigma^L_{\mu} \ell_{\nu} \Sigma^L_{\alpha} \ell_{\beta}
- i \Sigma^L_{\mu} \Sigma^L_{\nu} \Sigma^L_{\alpha} \ell_{\beta}
\right\rangle 
 - \left( L \leftrightarrow R \right) , 
 \label{eq:WZW2}
\ea
where
\be
\Sigma^L_\mu = U^{\dagger} \partial_\mu U \, , \qquad\qquad
\Sigma^R_\mu = U \partial_\mu U^{\dagger} \, ,
\label{eq:sima_l_r}
\ee
and
$\left( L \leftrightarrow R \right)$ stands for the interchanges
$U \leftrightarrow U^\dagger $, $\ell_\mu \leftrightarrow r_\mu $ and
$\Sigma^L_\mu \leftrightarrow \Sigma^R_\mu $.
The left and right sources, $\ell_\mu$ and $r_\mu$ respectively, are defined
in terms of the 
vector ($v_\mu$) and scalar ($a_\mu$) ones as
$
r_\mu = v_\mu + a_\mu, \ell_\mu = v_\mu - a_\mu$.
In the case of two flavours, singlet vector currents need to be included
for phenomenologically relevant processes. The Lagrangian remains the above
one but with the singlet vector field nonzero.
The two flavour case with an axial vector singlet as well is somewhat more
complicated and is discussed in Ref. \cite{kaiser}.

\setcounter{equation}{0}
\section{Renormalization}

In quantum field theory the corrections to the Born amplitude will lead
to unphysical 
ultra-violet (UV) divergences. Those
divergences should be 
at most polynomials in the external momenta and/or masses, thus all
non-analytical divergences
should cancel with each other. To define  
finite quantities one needs the inclusion of polynomial counter-terms that
render any observable 
free of UV divergences. In order to obtain the full structure of the needed
poles arising at  
one loop level in dimensional regularization we follow
the standard procedure and consider the fluctuations around the
classical solution of the equation of motion for the Goldstone boson
matrix 
$\bar{U}$
\begin{equation}
\label{eomS}
\frac{\delta S_2}{\delta U} = 0\quad  \Rightarrow \quad \bar{U}\,.
\end{equation}
The subindex in the
action functional refers in what follows to the chiral power.
This allows to write the expansion
\be
U= u(1 + i \xi -\frac{1}{2} \xi^2 + \ldots ) u\,, 
\ee
where $u^2(x) = \bar{U}(\pi(x))$ 
and $\xi(x)$ is a traceless hermitian matrix.
Eq.~(\ref{eomS}) defines the equation of motion (EOM) in terms of 
the covariant derivative of the $u(x)$ fields as
\be
\label{eom}
\nabla^\mu\um = 
\frac{i}{2}\left(\cm-\frac{1}{n}\trace{\cm}\right)\,,
\ee
where 
\begin{eqnarray}
u_\mu& =& i \left( u^\dagger \left(\partial_\mu - i r_\mu\right) u -
u \left(\partial_\mu - i \ell_\mu\right) u^\dagger \right)\,, \nonumber \\
\chi_\pm & = & u^\dagger \chi u^\dagger \pm  u \chi^\dagger u\,, 
\end{eqnarray}
and with  
\begin{equation}
\chi = 2 B_0 \left( s + i p \right)
\end{equation}
given in terms of the scalar and pseudoscalar sources $s$ and $p$ respectively.
$B_0$ is a constant not restricted by chiral symmetry and related with
the quark condendesate.
For any  operator $X$
the covariant derivative 
\begin{equation}
\nabla_\mu X = \partial_\mu X + \left[ \Gamma_\mu , X \right],
\end{equation}
is defined in terms of the connection
\begin{equation}
\Gamma_\mu = \frac{1}{2} \left( u^\dagger (\partial_\mu-i r_\mu) u + 
u(\partial_\mu-i \ell_\mu)u^\dagger \right)\,.
\end{equation}

We are interested in the second variation of the WZW action, $\xi^2$ terms.
Expanding the functional action up to this order 
\ba
S[U,{\bf j}] &=& S_2[\bar{U},{\bf j}] +  S[\bar{U},{\bf j}]_{\mathrm{WZW}} 
+ \Bigg( \frac{1}{2} \xi_i 
\frac{\delta S_2[\bar{U},{\bf j}]}{\delta\bar{U}_i(x_1)
\delta\bar{U}_j(x_2)} \xi_j
+  \frac{1}{2} \xi_i 
\frac{\delta S[\bar{U},{\bf j}]_{\mathrm{WZW}}}{\delta\bar{U}_i(x_1)
\delta\bar{U}_j(x_2)} \xi_j
\nonumber\\&&
+S_4[\bar{U},{\bf j}] 
+S_6^W[\bar{U},{\bf j}] \Bigg)
\ea
Here $S_2$ denotes the order $p^2$ chiral action, $S_4$ the order $p^4$,
and $S_6^W$ the $p^6$ one of odd intrinsic parity.
The first order derivative of $S_{\mathrm{WZW}}$ does not contribute to the
order we are considering.

It is easy to verify that the second variation of the integral over the
five-dimensional manifold, can be expressed as an integral over the ordinary
four-dimensional
spacetime (the integrand is a total derivative) \cite{DW,BBCzpc}. 
This gives us a hint about the
possible operators that can appear at the next chiral order. 
The result of the various terms combine in
a chirally covariant form,
\ba 
\label{2var}
\delta S_{\mathrm{WZW}} &=&
\frac{i N_c}{96 \pi^2} \int d^4x \epsilon^{\mu\nu\alpha\beta} \biggl\{
\langle \left( \xi {\nabla}_\mu \xi - {\nabla}_\mu \xi \xi \right) 
\left[ \frac{i}{8}
\un \ua \ub +\frac{1}{2}\un \fpab + \frac{1}{2} \fpna \ub\right] \rangle
\nonumber \\
&& +  \langle \left( \xi \um {\nabla}_\nu \xi - {\nabla}_\nu \xi
\um \xi  \right) \left[i \ua\ub +\frac{1}{2} \fpab \right] \rangle
\nonumber \\
&& +\frac{i}{4} \langle \xi \xi \left[ \um\un ,\fmab \right] \rangle 
-\frac{1}{8} \langle \xi \xi \left[ \fmmn , \fpab \right] \rangle \nonumber \\
&&-\frac{i}{4} \langle \xi \um \xi \left\{\un , \fmab \right\} \rangle
\biggr\},
\ea
where the operators
\begin{equation}
f_\pm^{\mu\nu} = u F_L^{\mu\nu} u^\dagger \pm u^\dagger F_R^{\mu\nu} u
\end{equation}
are defined in terms of the non-abelian field strengths
\begin{equation}
F_L^{\mu\nu} = \partial^\mu \ell^\nu - \partial^\nu \ell^\mu
-i[\ell^\mu,\ell^\nu]\,, \quad
F_R^{\mu\nu} = \partial^\mu r^\nu - \partial^\nu r^\mu
-i[r^\mu,r^\nu]\,.\nonumber
\end{equation}
It is now straightforward to compute in dimensional regularization, the
divergent part of the action in terms of an arbitrary  number of flavours
and colours   
\ba
\label{div}
Z_{\mathrm{1-loop}}^{\mathrm{WZ}\infty} &=& \frac{-1}{16 \pi^2 (d-4)}
\frac{N_c N_f}{1152 \pi^2 F^2} \Biggl\{ 4 O_1^W + 
\left(-3+\frac{6}{N_f^2} \right)
O_2^W - \frac{6}{N_f} O_3^W \nonumber \\
&&- 2 O_4^W + 4 O_5^W +\frac{8}{N_f} O_6^W 
+\left(-\frac{3}{2} + \frac{6}{N_f^2}
\right) O_{11}^W - 2 O_{12}^W -10 O_{13}^W  \nonumber \\
&& -3 O_{14}^W + O_{15}^W + 2 O_{16}^W + O_{17}^W + \frac{6}{N_f} O_{18}^W 
 \nonumber \\
&& - 4 O_{19^W} - O_{20}^W + 5 O_{21}^W + 4 O_{22}^W -\frac{6}{N_{f}} O_{24}^W
\Biggr\}\,.
\ea
The operators $O_i^W$ are defined in Table~\ref{tableNf}.
This agrees with the known result \cite{Akhoury,BBCzpc,Issler1}.
Notice however, that for the case $N_f=2$, the divergent part vanishes, due
to the Cayley-Hamilton relations. The reason for this is that in the $SU(2)$
case, there are no anomalies in the absence of a singlet vector source, i.e.
the homotopy group is reduced to the trivial element.
However, the physically interesting situation is when we have a singlet
vector source in the formalism \cite{kaiser}. In
this last case the initial symmetry is extended
to include electromagnetic effects. 
The initial symmetry group Eq.~(\ref{ssb})
is then modified to
\begin{equation}
\label{singletsym}
SU(2)_L \times SU(2)_R \times U(1)_V\,.
\end{equation}
Bearing in mind that the quark charge matrix,
$Q = {\mathrm{diag}}(2/3,-1/3)$, is not a generator of $SU(2)_L\times SU(2)_R$
the anomaly fails to vanish.
Allowing for such external sources, the divergences for the
two-flavour case are
\be
\label{divsin}
Z_{N_f=2}^{\mathrm{WZ}\infty} = \frac{-1}{16 \pi^2 (d-4)}
\frac{N_c}{1152 \pi^2 F^2} \Biggl\{
3\, o_6^W + 3\,o_7^W
 -\frac{3}{2}\,o_8^W + 6\, o_9^W -18\, o_{10}^W +12\, o_{11}^W
-12\, o_{13}^W
 \Biggr\}\,,
\ee
where the operators $o_i^W$ are listed in Table~\ref{tableNf:2}, and 
$o_6^W,\ldots,o_{13}^W$
represent the additional structures required by the inclusion of the singlet
vector source. Since the $p^4$ term only involves singlet currents in this
case, it is no surprise that the infinity can also be written in terms
of the extra operators only.

We used FORM~3 \cite{FORM} for some of the calculations in this section.
To ascertain the correctness of our results we have cross-checked the 
divergent parts of the $\eta \rightarrow \gamma \gamma \pi^0 \pi^0$ decay
in the case $N_f=3$ \cite{eta} and $\pi^0 \rightarrow \gamma \gamma$
and $\gamma \rightarrow
\pi \pi \pi$ for the $N_f=2$ case \cite{gammas}.

\setcounter{equation}{0}
\section{Next-to-Leading order Lagrangian for generic $N_f$}

In the previous section we have computed the divergent part of the 
WZW Lagrangian at next-to-leading order. 
As it has been mentioned already, for any phenomenological purpose, it is
crucial  
to work out a minimal  set of operators that reproduces the low-energy
dynamics. For this purpose the following list of building blocks are
sufficient at next-to-leading order
\be
\um\,, \quad \hmn = {\nabla}_\mu\un+{\nabla}_\nu\um\,,
\quad \fpmn\,, \quad \fmmn = {\nabla}_\nu\um-{\nabla}_\mu\un\,, 
\quad \cp\,, \quad \cm\,.
\ee
All others can be reduced to these.
The choice of basis is motivated first of all by the operators arising
in 
Eqs.~(\ref{div}), (\ref{divsin}), and second because of its relative simplicity
for reducing the number of terms.
For constructing the Lagrangian, besides the continuous symmetries, one has 
to implement the discrete symmetries. For the case of QCD: $P$ (parity), $C$
(charge
conjugation) and h.c. (hermiticity). The
transformation of the basic construction blocks under these
is in Table \ref{disc}. 
Hermitian conjugation merely determines
the presence of a global $i$ factor but the use of $C$ turned out to be
quite restrictive 
in combination with $P$.
In addition to this
we have in the anomalous sector the presence of
an $\epsilon^{\mu\nu\alpha\beta}$ tensor which is odd under parity. 

\begin{table}\begin{center}
\begin{tabular}{cccc}
\hline
\hspace{.5cm} operator \hspace{.5cm} & \hspace{1.5cm} $P$ \hspace{1.5cm}
 &\hspace{1cm} $C$ \hspace{1cm} & \hspace{1cm} h.c. \hspace{1cm} \\ \hline
 $u_\mu$ & $-\ve(\mu) u_\mu$ & $u_\mu^T$ & $u_\mu$ \\
 $h_{\mu\nu}$ & $-\ve(\mu)\ve(\nu) h_{\mu\nu}$ & $h_{\mu\nu}^T$ &
 $h_{\mu\nu}$ \\
 $\chi_\pm$ & $\pm \chi_\pm$ & $\chi_\pm^T$ & $\pm \chi_\pm$ \\
 $f_\pm^{\mu\nu}$ & $\pm \ve(\mu)\ve(\nu) f_\pm^{\mu\nu}$ &
 $\mp f_\pm^{\mu\nu T}$ & $f_\pm^{\mu\nu}$ \\
 \hline
 \end{tabular}
\end{center}
\caption{\label{disc}$P$, $C$ and
hermiticity properties of the basic operators.}
 \end{table}

We now sketch the arguments used during the construction:\\
i) If a derivative acting on $\chi_\pm$ appears
in a given 
operator, partial integration (PI) always allows to remove it.
The presence of $\epe$ allows for at most one power of $\chi_\pm$.\\
ii) Bianchi Identities (BI)
are used to remove all terms with more than two-derivatives. They
can all be rewritten into terms with $\fmmn$.\\
iii)
If only one
field strength is present in the operator, we can remove all the extra
derivatives acting on it by PI. \\
iv)
The EOM and commutators allow to remove $\nabla^\mu\hmn$.
Using in addition
commutators and antisymmetry properties of the indices, we can always remove
other terms with an extra derivative on a $h$.\\
v)
Terms involving the external fields only for the $N_f$-flavour case
cannot be constructed. Terms with one $\chi$ or $\chi^\dagger$
are obviously not chiral invariant,
terms with two field strengths and two extra derivatives
can always be related to terms with 3 field strengths using partial
integration and the BI. And finally, terms with 3 field strengths are forbidden
since the $C$ and $P$ transformations clash.

Using these rules the full list of operators for the
$N_f$-flavours case contains 57 monomials.
Further reduction
requires a more extensive study of partial integrations and the use
of the lowest order EOM, Eq.~(\ref{eom}),
the Bianchi identities and the Schouten identity \cite{Schouten}.

The use of the EOM is equivalent, at lowest order, to a field redefinition,
because 
the generating functional at ${\cal O}(p^6)$ contains the classical
Lagrangian density at ${\cal O}(p^6)$, see the proof in \cite{BCGlag}
as well as the discussion in \cite{FearingScherer1}.

The BI yield two relations.
The first one follows from the BI of the field strength tensor
$\Gamma_{\mu\nu}$ 
\begin{equation}
\nabla_\mu \Gamma_{\nu\rho} + \nabla_\nu \Gamma_{\rho\nu} + 
\nabla_\rho \Gamma_{\nu\mu} =
0\,, 
\end{equation}
and reads
\be
\nabla_\mu\fpna+\nabla_\nu f_{+\alpha\mu}+\nabla_\alpha\fpmn
= \frac{i}{2}\left(
\com{\fmmn,\ua}+\com{\fmna,\um}+\com{f_{-\alpha\mu},\un}\right)\,,
\ee
while the second arises using the identity
\begin{equation}
f_-^{\mu\nu} = \nabla^\nu u^\mu - \nabla^\mu u^\nu\,,
\end{equation}
which leads to
\be
\nabla_\mu\fmna+\nabla_\nu f_{-\alpha\mu}+\nabla_\alpha\fmmn
= \frac{i}{2}\left(
\com{\fpmn,\ua}+\com{\fpna,\um}+\com{f_{+\alpha\mu},\un}\right)\,.
\ee
We shall also apply the Schouten identity \cite{Schouten}, 
\be
A^\gamma\epe
-A^\mu\epsilon^{\gamma\nu\alpha\beta}
-A^\nu\epsilon^{\mu\gamma\alpha\beta}
-A^\alpha\epsilon^{\mu\nu\gamma\beta}
-A^\beta\epsilon^{\mu\nu\alpha\gamma}=0\,,
\ee
which holds for any operator A and can in principle be applied twice to the
terms with $6$ indices.

In the generic $SU(N_f)_L\times SU(N_f)_R$
case, without the inclusion of a vector singlet field, 
there are only 33 linearly independent relations which follow from the use of PI, BI and
Schouten identities. This leaves us with 24
independent monomials contributing to the
$N_f$-flavour Lagrangian 
\be
{\cal L}_6^W = \sum_{i=1}^{24} K_i^W O^W_i,
\ee
where the $K_i^W$ are coupling constants. The divergences 
can be subtracted using the usual modified $\overline{\mathrm{MS}}$ scheme
\be
K_i^W =  K_i^{Wr} + \eta_i^{(N_f)} \frac{\mu^{d-4}}{16\pi^2}
\left\{\frac{1}{d-4}-\frac{1}{2}\left(\ln(4\pi)+\gamma+1\right)\right\}\,, 
\ee
where the subtraction
coefficients $\eta_i^{(N_f)}$ can be read from Eq.~(\ref{div}).
The operators $O^W_i$ 
are listed in Table \ref{tableNf} using the notation
\begin{equation}
\com{a,b,c} = abc-cba \quad \acom{a,b,c} = abc +cba\,.
\end{equation}
The subtraction coefficients are repeated there for completeness.
\begin{table}\begin{center}
\renewcommand{\arraystretch}{1.2}
\begin{tabular}{lcccc}
\hline
\hspace{1cm} monomial ($O_i^W$)  & $i$ $N_f$-flavours &  $
384 \pi^2 F^2 \eta_i^{(N_f)} $& $i$ 3-flavours &$384 \pi^2 F^2 \eta_i^{(3)}$ \\
\hline
$ i\epe\trace{\cm\um\un\ua\ub}$        & 1 &$4 N_f$   & 1 & $12$
\\
$\epe\trace{\cp\com{\fmmn,\ua\ub}}$    & 2  &$\left( -3 N_f +
\frac{6}{N_f}\right)$                                 &2 &  $-7$
\\
$\epe\trace{\cp\um}\trace{\un\fmab}$    & 3  &$-6$    &3 & $-6$
\\
$\epe\trace{\cm\acom{\fpmn,\ua\ub}}$    & 4  &$-2 N_f$&4 & $-6$
\\
$\epe\trace{\cm\um\fpna\ub}$            & 5  &$4 N_f$ &5 & $12$
\\
$\epe \trace{\cm}\trace{\fpmn\ua\ub}$   & 6  &$8$  &6 &   $8$
\\
$i\epe\trace{\cm\fpmn\fpab}$            & 7  &$0$  &7 &  $0$ 
\\
$i\epe\trace{\cm}\trace{\fpmn\fpab}$    & 8  &$0$  &8 &  $0$
\\
$i\epe\trace{\cm\fmmn\fmab}$            & 9  &$0$  &9 &  $0$ 
\\
$i\epe\trace{\cm}\trace{\fmmn\fmab}$    & 10 &$0$  &10 &  $0$
\\
$i\epe\trace{\cp\com{\fpmn,\fmab}}$     & 11 &$\left( -\frac{3 N_f}{2} +
\frac{6}{N_f} \right)$                             &11  &$-\frac{5}{2}$
\\
$\epe\trace{\hgm\com{\ug,\un\ua\ub}}$   & 12  &$-2 N_f$ &12 & $-6$
\\
$i\epe\trace{\hgm\acom{\fpgn,\ua\ub}}$  & 13  &$- 10 N_f$  &13 &$-30$
\\
$i\epe\trace{\hgm\com{\fpna,\ug,\ub}}$  & 14  &$-3 N_f$  &14 &$-9$
\\
 $i\epe\trace{\hgm\com{\ug,\fpna,\ub}}$ & 15  &$N_f$  &15 & $3$
\\
$\epe\trace{\fmgm\com{\ug,\un\ua\ub}}$  & 16  &$2 N_f$  &16 &$18$
\\
$\epe\trace{\fmmn\com{\ug\ug,\ua\ub}}$  & 17  &$N_f$  &17 & $15$
\\
$\epe\trace{\fmmn\ua}\trace{\ug\ug\ub}$ & 18  &$6$    &18 & $18$
\\
$i\epe\trace{\fpgm\acom{\fmgn,\ua\ub}}$ & 19  &$-4 N_f$&19 &$-12$
\\
$i\epe\trace{\fpgm\com{\fmna,\ug,\ub}}$ & 20  &$-N_f$  &20 & $-3$
\\
$i\epe\trace{\fpgm\com{\ub,\fmna,\ug}}$ & 21  &$5 N_f$  &21 & $15$
\\
$\epe\trace{\um\acom{\nabla_\gamma\fpgn,\fpab}}$&22&$4 N_f$  &22 &$12$
\\
$\epe\trace{\um\acom{\nabla_\gamma\fmgn,\fmab}}$&23&$0$  &23 &$0$
\\
$\epe\trace{\fmmn\acom{\ug,\ua}}\trace{\ug\ub}$ & 24&$-6$  &-  & -
\\
\hline
 \end{tabular}
\end{center}
\caption{
\label{tableNf}
The list of operators $O_i^W$
of the $p^6$ odd intrinsic parity or anomalous chiral Lagrangian.
For $N_f$ flavours all 24 are relevant, for 3 flavours $O_{24}^W$
can be dropped. The renormalization coefficients $\eta_i^{N_f}$
are listed as well.
The flavour trace is denoted by
$\lgl \dots \rgl$ and we use the notation 
$\com{a,b,c} = abc-cba$ and $\acom{a,b,c} = abc +cba$.
}
\end{table}

\setcounter{equation}{0}
\section{A minimal set for the Lagrangian with $N_f=2,3$}

Besides the previous arguments one can make use of the Cayley-Hamilton (CH)
relation to reduce the number of  
operators for the 3 and 2-flavour case. Those relations follow directly
from the characteristic  
polynomial of any matrix. Their use is rather common and we refer to any
basic text book in linear algebra or to \cite{FearingScherer2,BCGlag}.

In the three flavour case the use of the CH includes only one additional
relation with respect to the generic case
\begin{equation}
\label{ch3}
0\, =\, 2 O_{16}^W\, +\, 2 O_{17}^W\, +\, 2\, O_{18}^W\,+\,O_{24}^W\,.
\end{equation}
Thus the number of independent operators for the case $N_f=3$ is 23.
They are listed 
in Table~\ref{tableNf} and they form 
the Lagrangian density 
\begin{equation}
{\cal L}_6^W = \sum_{i=1}^{23} C_i^W O^W_i\,.
\end{equation}

As in the 
previous case the infinities can be subtracted using
\be
C_i^W =  C_i^{Wr} + \eta_i^{(3)} \frac{\mu^{d-4}}{16\pi^2}
\left\{\frac{1}{d-4}-\frac{1}{2}\left(\ln(4\pi)+\gamma+1\right)\right\}\,
\ee
where the coefficients $\eta_i^{(3)}$ can be read directly from Eq.~(\ref{div}) after
using the additional relation (\ref{ch3}).

In the two flavour case, the CH relations turn out to be very constraining
and the list of 24 monomials in Table~\ref{tableNf} can be reduced up to 5.
These are the first five operators $o_1^W,\ldots,o_5^W$
listed in Table \ref{tableNf:2}.
In addition to those operators the physically interesting case must include the
singlet vector source, which is not present in the $SU(N_f)_L\times SU(N_f)_R$
formalism [c.f. (\ref{singletsym})] thus allowing a  $U(1)$ operator
with nonzero trace, 
i.e. $\langle f_+^{\mu\nu} \rangle \ne 0$. 
In this last case, symmetry requirements allow the construction of 
an additional 
set of 28 monomials. The use of PI, BI and CH reduces 
the additional terms up to 8
linear independent operators.
The Lagrangian in this case is given by
\be
{\cal L}_6^W = \sum_{i=1}^{13} c_i^W o^W_i\,,
\ee
where the minimal list of operators $o_i$ 
is given in Table \ref{tableNf:2}. 

As in the general case the infinities can be subtracted using
\be
c_i^W =  c_i^{Wr} + \eta_i^{(2)} \frac{\mu^{d-4}}{16\pi^2}
\left\{\frac{1}{d-4}-\frac{1}{2}\left(\ln(4\pi)+\gamma+1\right)\right\}\,, 
\ee
where the $ \eta_i^{(2)}$ coefficients are listed in Table~\ref{tableNf:2},
they follow from
Eq.~(\ref{2var}) and the use of relations between these operators
when we restrict to the $SU(2)$ algebra and directly from Eq.~(\ref{divsin}).

\begin{table}\begin{center}
\renewcommand{\arraystretch}{1.2}
\begin{tabular}{lcc}
\hline
\hspace{2cm} monomial ($o_i^W$)  & $i$ 2-flavour & $384
\pi^2 F^2 \eta_i^{(2)} $ \\
\hline
$\epe\trace{\cp\com{\fmmn,\ua\ub}}$    & 1 & 0
\\
$\epe\trace{\cm\acom{\fpmn,\ua\ub}}$    & 2 & 0
\\
$i\epe\trace{\cm\fpmn\fpab}$            & 3 & 0
\\
$i\epe\trace{\cm\fmmn\fmab}$            & 4 & 0
\\
$i\epe\trace{\cp\com{\fpmn,\fmab}}$     & 5 & 0
\\
\hline
$\epe \trace{\fpmn} \trace{\cm\ua\ub}$  & 6 & 3
\\
$ i \epe \trace{\fpmn} \trace{\fpab\cm}$ & 7 & 3
\\
$ i \epe \trace{\fpmn} \trace{\fpab}\trace{\cm} $ & 8 & $-\frac{3}{2}$
\\
$ i \epe \trace{\fpgm} \trace{\hgn\ua\ub}$ & 9 & 6
\\
$ \displaystyle i\epe \trace{\fpgm} \trace{\fmgn\ua\ub}$ & 10 & $-18$
\\
$ \epe \trace{\fpmn} \trace{\fpga\hgb} $ & 11 & 12
\\
$  \epe \trace{\fpmn} \trace{\fpga\fmgb}$ & 12 & 0
\\
$\epe \trace{\nabla_\gamma \fpgm} \trace{\fpna\ub}$ & 13 & $-12$
\\
 \hline
 \end{tabular}
\end{center}
\caption{
\label{tableNf:2}
The operators $o_i^W$ and the divergent piece $\eta_i^{(2)}$. 
Notation as in Table~\ref{tableNf}. The first five are for traceless
external currents. The last 8 are needed if a singlet external vector current
is present.}
 \end{table}

\setcounter{equation}{0}
\section{Conclusions}

In this work we have determined the minimal anomalous chiral Lagrangian
at order $p^6$. The minimal set of terms is 24 in the general flavour case,
23 for three flavours and 13 for the two-flavour case including for the latter
the singlet vector field since it is physically relevant.
We have also recalculated the infinite parts for the general flavour case
which were known earlier \cite{Akhoury,BBCzpc,Issler1} and 
determined the infinities
in the two-flavour case in the presence of a singlet vector field.

\vskip.4cm
\noindent {\bf Acknowledgements}\\
PT was supported by the EU network EURODAPHNE, EC--Contract No.
ERBFMRX--CT980169. LG acknowledges partial
support from European Program HPRN-CT-2000-00149.
JB thanks the Institute for Nuclear Theory at the
University of Washington for hospitality and the Department of Energy for
partial support during the completion of this work.
We thank B.~Moussallam for pointing out an inconsistency in the
infinities in Eq. (\ref{divsin}) in the first version.

\vskip.4cm
\noindent {\it Note added }\\
When completing this work, Ref.~\cite{Fearing2} appeared.
The authors work in a different basis but we fully agree on the number
of terms needed. Our notation was chosen to match the one of
\cite{BCGlag}.

\end{document}